\documentstyle[prd,aps, eqsecnum]{revtex}
\begin{document}
\draft

\title{Conformal Black Hole Solutions of  Axi-Dilaton Gravity in D-dimensions}
\author{ H. Cebeci ,
T. Dereli \\ {\small Department of Physics, Middle East Technical
University} \\ {\small 06531 Ankara, Turkey}}

\maketitle

\bigskip
\begin{abstract}

\noindent Static, spherically symmetric solutions of 
axi-dilaton gravity in $D$ dimensions is given in the Brans-Dicke frame
for arbitrary values of the Brans-Dicke constant $\omega$ and 
an axion-dilaton coupling parameter $k$.
The mass and the dilaton and axion charges are determined
and a BPS bound is derived. 
There exists a one parameter family of 
black hole  solutions in the scale invariant limit.
\end{abstract}

\bigskip

\section{Introduction}

It is an exciting conjecture that all superstring models belong to 
a  hypothetical eleven dimensional M-theory that would accommodate
their apparent dualities.
 M-theory as a classical theory may be considered in a
low-energy limit where only the low-lying massless excitation modes contribute 
to  an effective field  theory. As such it would be the same as 
the eleven dimensional simple supergravity theory.
A subsequent Kaluza-Klein reduction would bring it to a 
ten dimensional theory related  with the  type IIA
string model whose gravitational sector
consists of the space-time metric tensor $g$, dilaton scalar
$\phi$ and the axion potential $(p+1)$-form $A$  that
would minimally couple to $p$-branes.
We call such an effective gravitational field theory 
an axi-dilaton gravity in $D$-dimensions and consider in the following 
its static, spherically symmetric solutions for $p=D-4$.

The study of black hole solutions of higher dimensional gravity
theories started in 1963 with the generalisation of Schwarzschild and
Reissner-Nordstr\"{o}m solutions to $D>4$ dimensions
by Tangherlini \cite{tangherlini}.
These solutions were later put in a wider context by Myers and Perry
\cite{myers}, while Gibbons and Maeda \cite{gibbons}
emphasised the relevance of dilaton scalars
for the interpretation of such solutions.
They provided a wide range of static, spherically symmetric solutions
of the coupled Einstein-antisymmetric tensor-massless scalar field
equations (see also \cite{xantha1},\cite{kim}). 
On the other hand 
it is a well-known fact that scalar-tensor
theory of Brans-Dicke \cite{brans and dicke} may be re-written in
terms of 
a conformally re-scaled metric as the coupled Einstein-massless scalar
field theory \cite{deser}, \cite{anderson}, \cite{parker}.
For a particular value of the Brans-Dicke coupling parameter, namely
for $\omega = -\frac{3}{2}$ in $4$-dimensions,  theory
becomes locally scale invariant and called the Einstein-conformal scalar
field theory. 
We had shown in a previous work that
the  conformal re-scaling properties of the Brans-Dicke theory can be
conveniently exhibited using the non-Riemannian reformulation
involving space-time torsion
expressed in terms of the gradient of the scalar field \cite{dereli1}.
Brans-Dicke theory has also been generalised to $D$-dimensions \cite{dereli2}
and the black hole solutions of the Brans-Dicke-Maxwell field
equations were given \cite{cai},\cite{dereli3}.

In a remarkable paper Bekenstein \cite{bekenstein1} has found two 
classes of static,
spherically symmetric solutions of the Einstein-conformal scalar field
equations, and argued \cite{bekenstein2} that one particular class describes
black hole solutions with scalar hair.
His arguments were later repeated in  $D>4$ dimensions \cite{xantha2}.
It is essential here to note that such a  subclass of conformal black 
hole solutions cannot be reached by the assumptions of Ref.\cite{gibbons}.
In this paper  we consider  axi-dilaton gravity 
in $D$-dimensions ($p=D-4$) in the Brans-Dicke frame
and give its  static, spherically symmetric solutions
for arbitrary values of two  coupling parameters $\omega$ and $k$. 
A one-parameter family of conformal black hole solutions are obtained
for $\omega = -\frac{D-1}{D-2}$ and $k =-\frac{D-4}{D-2}$. 

\section{Axi-dilaton Gravity in $D$-dimensions}

The dynamics of the axi-dilaton gravity 
will be determined by a variational principle from the
 action $ I[e, \omega, \phi, A] = \int {\cal{L}}$
 where the Lagrangian density $D$-form is taken
in the Brans-Dicke frame as
\begin{equation}
{\cal{L}}=\frac{\phi}{2} R_{ab} \wedge \ast(e^{a} \wedge e^{b})-
\frac{\omega}{2\phi}
d\phi \wedge \ast d\phi - \frac{\phi^k}{2} H \wedge \ast H\ .
\end{equation}
Here the basic gravitational field variables are the co-frame 1-forms
$e^{a} $ , in terms of 
which the space-time
metric {\normalsize \ }$g=\eta _{ab}e^{a}\otimes e^{b}$ where
$\eta_{ab} = diag(-+++ \dots )$.
The Hodge $\ast$ map is defined 
so that the oriented volume form
$\ast 1 = e^0 \wedge e^1 \wedge \dots e^n.$
The metric compatible, torsion-free  connection 1-forms
$\omega _{\;\ b}^{a}$ are obtained from the Cartan structure equations
\begin{equation}
de^{a}+\omega _{\;\ b}^{a}\wedge e^{b} = 0,
\end{equation}
and the corresponding curvature 2-forms 
\begin{equation}
R_{\;\ b}^{a}=d\omega_{\;\ b}^{a}+\omega _{\;\ c}^{a}\wedge \omega_{\;\ b} ^{c}. \quad
\end{equation}
$\phi \ $ is the  dilaton 0-form and $H$ is a $\left( p+2\right) $-form 
field that is derived from the $\left( p+1\right) $-form axion
potential $A$ such that $H=dA$.  
$ \omega $
and $k$ are real parameters.

The  field equations obtained 
from this action are 
\begin{equation}
-\frac{\phi}{2}R^{bc}\wedge {}^\ast (e_{a}\wedge e_{b}\wedge e_{c}) = 
\frac{\omega}{\phi} \tau_{a}[\phi] + \phi^{k} \tau_{a}[H] 
+ D(\iota_{a} \ast d\phi)
\quad ,
\end{equation}
\begin{equation}
\tilde{k} d\ast d\phi = \frac{\alpha}{2} \phi^{k} H\wedge \ast H \quad ,
\end{equation}
\begin{equation}
d(\phi^{k} \ast H ) = 0 \quad, \quad dH = 0 ,
\end{equation}
where the dilaton and axion stress-energy $(D-1)$-forms are given by
\begin{equation}
\tau_{a}[\phi] = 
\frac{1}{2} (\iota_{a}d\phi \wedge \ast d\phi +
d\phi \wedge \iota_{a} \ast d\phi ) 
\end{equation} 
and
\begin{equation}
\tau_{a}[H] = \frac{1}{2} (\iota_{a}H\wedge \ast H+ (-1)^{p-1}H \wedge
\iota_{a} \ast H ) ,
\end{equation}
respectively. We set $\alpha = \frac{2p-(n-3)}{n-1} + k$ and
$\tilde{k} = \frac{n}{n-1} + \omega.$

The same action may be rewritten in terms of the 
$(D-p-2)$-form field
\begin{equation}
G \equiv \phi^{k} \ast H
\end{equation}
that is dual to the axion $(p+2)$-form field $H$. 
We have in terms of $G$,
\begin{equation}
{\cal{L}} = \frac{\phi}{2}R_{ab}\wedge \ast \left( e^{a}\wedge e^{b}\right) 
- \frac{\omega}{2\phi}d\phi \wedge \ast d\phi + \frac{\phi^{-k}}{2}
G \wedge \ast G .
\end{equation}
Hence given any solution $\{g, \phi, H\}$ of the field equations 
derived from (1), we may write down a dual 
solution $\{g, \phi, G \}$
to the field equations derived from (10). This 
notion of duality generalises the
usual electric-magnetic duality in $D=4$ source-free electromagnetism.  

Finally we wish to point out that the passage to the Einstein frame is achieved
by the following conformal re-scaling of the field variables:
\begin{eqnarray}
\tilde{g} &=& \phi^{\frac{2}{(n-1)}} g  \nonumber \\
\tilde{\phi} &=& {\tilde{k}}^{1/2} \ln \phi   \nonumber \\
\tilde{H} &=& H .
\end{eqnarray}
The resulting Lagrangian density D-form will be
\begin{equation}
{\cal{L}} = 
\frac{1}{2}{\tilde{R}}_{ab}\wedge \tilde{\ast} \left({\tilde{e}}^{a}\wedge 
{\tilde{e}}^{b}\right) 
- \frac{1}{2}d{\tilde{\phi}} \wedge \tilde{\ast} d{\tilde{\phi}} - \frac{1}{2}
exp(\frac{\alpha}{{\tilde{k}}^{1/2}} \tilde{\phi})
\tilde{H} \wedge \tilde{\ast}\tilde{H} .
\end{equation}
Given the above information, it is not difficult to compare solutions 
obtained in the Brans-Dicke frame with those given in the Einstein frame.

\section{Static, Spherically Symmetric  Solutions} 

We will be giving below  the most general
static, spherically symmetric $p = (D-4)$
brane solution to the field equations (2.4)-(2.6).
This family of  solutions generalises the usual 
magnetically charged Reissner-Nordstr\"{o}m black hole solution in $D =4$ 
to higher dimensions in a natural way.
To this end we start with the ansatz
\begin{equation}
g = -f^{2}(r) dt\otimes dt + h^{2}(r) dr\otimes dr + R^{2}(r) d\Omega _{n-1}
\end{equation}
for the metric tensor ($D=n+1$), 
$$\phi = \phi(r)$$  for the dilaton 0-form and 
$$H = g(r) e^{1}\wedge e^{2}\wedge e^{3}...\wedge e^{n-1}$$
for the axion field $(D-2)$-form.
We set  $ e^0 = f(r) dt$ and $e^n = h(r) dr$.
Then the Einstein field equations 
reduce to the following set of ordinary coupled differerential equations:
(${}^\prime$ denotes partial derivative with respect to  $r$)
\begin{eqnarray}
\phi \left [ \frac{(n-2)(n-1) h}{2 R^2} \left ( 1- (\frac{R^{\prime}}{h})^{2}
\right)
 -\frac{(n-1)}{R} ( \frac{R^{\prime}}{h})^{\prime} \right ] \nonumber \\
= \frac{\omega}{2\phi} (\frac{{\phi^{\prime}}^2}{h}) 
+\frac{\phi^k}{2} g^{2}h  + ( \frac{\phi^{\prime}}{h})^{\prime} + 
(n-1)\frac{\phi^{\prime} R^{\prime}}{h R} ,
\end{eqnarray}
\begin{eqnarray}
\phi \left [ \frac{(n-2) f^{\prime} R^{\prime}}{hR} 
+ ( \frac{f^{\prime}}{h})^{\prime} 
+ \frac{(n-2) f}{R} (\frac{R^{\prime}}{h})^{\prime}
- \frac{(n-3)(n-2) f h}{2 R^2} ( 1-(\frac{R^{\prime}}{h})^{2} )
\right ] \nonumber \\
=  -\frac{\omega f}{2\phi h} {\phi^{\prime}}^2  + \frac{\phi^k g^{2}fh}{2} 
- (\frac{\phi^{\prime} f}{h})^{\prime} - (n-2) \frac{\phi^{\prime} f}{h R} , 
\end{eqnarray}
\begin{eqnarray}
\phi \left [\frac{(n-1)(n-2)f}{2 R^2}   
\left (1- (\frac{R^{\prime} }{h})^{2} \right ) - 
(n-1)\frac{f^{\prime}R^{\prime}}{h^{2}R} \right ] \nonumber \\
= -\frac{\omega f {\phi^{\prime}}^2}{2 \phi h^2}
+ \frac{g^{2}f \phi^{k}}{2} + \frac{f^{\prime} \phi^{\prime}}{h^2}
+ (n-1)  \frac{f R^{\prime} \phi^{\prime}}{h^2 R} ;
\end{eqnarray}
while the dilaton field equation becomes
\begin{equation}
\tilde{k} {({\phi^\prime} \frac{f}{h} R^{n-1})}^{\prime}
= \frac{\alpha}{2} \phi^k g^{2} f h R^{n-1}
\end{equation}
and the axion field equation reads
\begin{equation}
(gR^{n-1})^{\prime} = 0 .
\end{equation}

Solutions to the above field equations can be written as:
\begin{equation}
\begin{array}{c}
R(r) = r \left ( 1-\left( \frac{C_{1}}{r}\right)^{n-2}\right )^{\alpha_{3}} \\  \\ 
f(r) = \left ( 1-\left( \frac{C_{2}}{r}\right)^{n-2}\right )
^{\alpha_4}\left ( 1-\left( \frac{C_{1}}{r}\right)^{n-2}\right )
^{\alpha _5} \\  \\
h(r) = \left ( 1-\left( \frac{C_{2}}{r}\right )^{n-2}\right )^{\alpha_2} 
\left ( 1-\left( \frac{C_{1}}{r}\right ) ^{n-2}\right )^{\alpha _{1}} \\ \\
\phi = \left (1-(\frac{C_{1}}{r})^{n-2}\right )^{\frac{2\gamma}{\alpha }}
\\  \\
g(r) = \frac{Q}{R^{n-1}}
\end{array}
\end{equation}
where $C_{1}$ and $C_{2}$ are two independent
integration constants and 
the third integration constant
$$Q^{2}=\frac{4\tilde{k} \gamma(C_{1}C_{2})^{n-2}(n-2)^{2}}{\alpha^{2}}.$$
The exponents are 

$$\alpha_{1} =\gamma(\frac{1}{(n-2)}-\frac{2}{(n-1)\alpha}) - \frac{1}{2}
\quad , \quad 
\alpha_{2} = -\frac{1}{2}  \quad , $$
$$\alpha_{3} = \gamma(\frac{1}{(n-2)}-\frac{2}{(n-1)\alpha})   \quad , $$ 

$$\alpha_{4} = \frac{1}{2}   \quad ,  \quad
\alpha_{5} = -\gamma(1 + \frac{2}{(n-1)\alpha}) +\frac{1}{2}   
\quad , $$  
with 
$$\gamma = \frac{\alpha^{2}(n-1)}{4 \tilde{k}(n-2)+\alpha
^{2}(n-1) } \quad .$$

\bigskip

\noindent  Some special cases deserve attention: 
 
\noindent   i)  For $Q=0$ and $\phi = const.$, 
we obtain the Tangherlini solution \cite{tangherlini} which is the generalisation of the
Schwarzschild solution in $D=n+1$ dimensions,
\begin{equation}
g=-\left( 1-\frac{2M}{r^{n-2}}\right) dt^{2}+\left( 1-\frac{2M}{r^{n-2}}%
\right) ^{-1}dr^{2}+r^{2}d\Omega _{n-1}
\end{equation}

\noindent ii) For $k = 0$ and $\phi = const. $,
we obtain the $D=n+1$-dimensional generalisation of the 
Reissner-Nordstr\"{o}m metric 
\begin{equation}
 \begin{array}{c}
g=-\left ( 1+\frac{Q^{2}}{(n-1)(n-2)r^{2(n-2)}}-\frac{2M}{r^{n-2}}\right )
dt^{2} \\ 
+\left ( 1+\frac{Q^{2}}{(n-1)(n-2)r^{2(n-2)}}-\frac{2M}{r^{n-2}}\right )^{-1}
dr^{2}+r^{2}d\Omega _{n-1} .
\end{array}
\end{equation}
The electric dual of this solution was also given by Tangherlini.

\noindent iii) For $Q=0$ we obtain solutions that generalise the 
Janis-Newman-Winicour solutions of the
Einstein-massless scalar field equations to 
$D$ dimensions \cite{xantha1}: 
\begin{equation}
\begin{array}{c}
R(r) = r h(r) \\ 
f(r) = \left ( \frac{r^{n-2}-r_{0}^{n-2}}{r^{n-2}+r_{0}^{n-2}}\right )
^{\beta _{1}-\beta_{2}} \\ 
h(r) = \left ( 1-\left( \frac{r_{0}}{r}\right) ^{2(n-2)}\right )
^{1/(n-2)}\left ( \frac{r^{n-2}-r_{0}^{n-2}}{r^{n-2}+r_{0}^{n-2}}\right )
^{-\beta_{1}/(n-2)-\beta_{2}} \\ 
\phi(r) = \left ( \frac{r^{n-2}-r_{0}^{n-2}}{r^{n-2}+r_{0}^{n-2}}\right )^{\beta_2}
\end{array}
\end{equation}
where in order to ease comparison we use the 
parametrisation
$$ \beta_{2} =\sqrt{\frac{4(n-1)}{(n-2) \tilde {k}}(4-{\beta_1}^2)}$$
and $\beta _{1}$ satisfies $4(n-2)r_{0}^{n-2}\beta_{1}=C , \quad $ $r_{0}$ 
and $ C$ being integration constants.

A consideration of the asymptotic behaviour of the fields in Brans-Dicke frame
will allow us to
determine a relationship satisfied by the mass, dilaton charge and magnetic
charge $Q$. The mass of the black hole is defined to be
\begin{equation}
2M \equiv \lim_{r\rightarrow \infty } r^{n-2}( 1 - f^{2}) = (C_{2})^{n-2}+ 
( \tilde{\gamma}- 2 \gamma ) (C_{1})^{n-2}
\end{equation}
where $\tilde{\gamma} = 1 - \frac{4\gamma}{(n-1)\alpha}$.
The scalar charge 
\begin{equation}
\Sigma \equiv \lim_{r \rightarrow \infty } r^{n-1}\frac{\phi^\prime}{\phi} 
= 2(n-2) \frac{\gamma }{\alpha }(C_{1})^{n-2}.
\end{equation}
Finally the magnetic charge can be found from
\begin{equation}
Q \equiv \lim_{r\rightarrow \infty } r^{n-1}g = Q .
\end{equation}
Therefore, by eliminating the integration constants $C_{1}$ and $C_{2}$
above we can find the following relationship between these 3 physical parameters:
\begin{equation}
Q^{2} = \frac{2(n-2)\Sigma }{\alpha} \tilde{k} 
\left [ \left(2\gamma -\tilde{\gamma}\right) \frac{\alpha \Sigma}
{2(n-2) \gamma}+ 2M \right ] \quad .
\end{equation}
From this relationship, since $\Sigma$ is a real parameter, the BPS bound
respected by the mass and charge 
of a black hole follows:

\begin{equation}
M \geq \frac{1}{2(n-2)}
\sqrt{\frac{4 (n-2)\tilde{k}+4 \alpha -{\alpha} ^{2}
(n-1)}{(n-1) \tilde{k}}} \, |Q| 
\end{equation}
provided
\begin{equation}
{\alpha} ^{2} (n-1)-4 \alpha \leq 4(n-2) \tilde{k}.
\end{equation}

\section{Conclusion}

A conformally scale invariant theory 
is obtained for the parameter values
$\omega = -\frac{n}{(n-1)}$ and $k = - \frac{(n-3)}{(n-1)}$.
A class of static, spherically symmetric solutions to the
conformally scale invariant theory 
may be reached from the solutions $(3.7)$ above by taking the limit
$ \alpha \rightarrow 0$ and $\tilde{k} \rightarrow 0$  
with the ratio $ \frac{2\gamma}{\alpha}$ kept fixed:
\begin{equation}
\begin{array}{c}
R(r) = r \left ( 1-\left( \frac{C_{1}}{r}\right)^{n-2}\right )^{-\frac{\beta}{(n-1)}} \\  \\ 
f(r) = \left ( 1-\left( \frac{C_{2}}{r}\right)^{n-2}\right )
^{1/2} \left ( 1-\left( \frac{C_{1}}{r}\right)^{n-2}\right )
^{1/2 - \frac{\beta}{(n-1)}} \\  \\
h(r) = \left ( 1-\left( \frac{C_{2}}{r}\right )^{n-2}\right )^{-1/2} 
\left ( 1-\left( \frac{C_{1}}{r}\right ) ^{n-2}\right )^{-1/2 - \frac{\beta}{(n-1)}} \\ \\
\phi = \left (1-(\frac{C_{1}}{r})^{n-2}\right )^{\beta} \\  \\
g(r) = \frac{Q}{R^{n-1}}
\end{array}
\end{equation}
where $C_1$ and $C_2$ are constants and $\beta$ and $Q$ should satisfy
$$ 2\beta (n-2)^{2}(C_1 C_2)^{n-2} = Q^2 .$$

The special case  of parameter values $Q=0$ and $C_2 =0$ in $D=4$
dimensions brings (4.1)  to
Bekenstein's Einstein-conformal scalar solution \cite{bekenstein1}.
The fact that this solution describes black holes was 
later clarified by Bekenstein \cite{bekenstein2}. His argument is based on the
observation that  the scalar particles being  postulated to
follow geodesic world-lines in Brans-Dicke theory \cite{dicke}
pre-supposes a particular type of scalar field coupling to matter.
On the other hand by assuming a different type of matter couplings 
one can show that neutral test particles would follow
conformal world-lines as argued by G\"{u}rsey \cite{gursey}
and Dirac \cite{dirac}. This assumption implies in particular 
that solution (4.1) describes   a black hole with  finite scalar charge. 
In a recent paper it was shown  that the conformal world-lines  
are nothing but autoparallel
curves in the non-Riemannian re-formulation of the Brans-Dicke theory \cite{dereli4}.
A further re-evaluation of 
the locally scale invariant solutions above  from a non-Riemannian point of
view  will be taken up in a separate study.

\section{Acknowledgement}

We are grateful to T\"{U}B\.{I}TAK (Scientific and Technical Research
Council of Turkey)
for partial support through  BAYG-BDP program.

\end{document}